\documentclass[aip,reprint]{revtex4-1}
\usepackage{graphicx}% Include figure files
\usepackage{dcolumn}% Align table columns on decimal point
\usepackage{bm}% bold math
\usepackage{CJK}

\begin{document}

%\preprint{}

\title{A systematic study of Rayleigh-Brillouin scattering in air, N$_2$ and O$_2$ gases}


\author{Ziyu Gu}
%\thanks{}
\affiliation{Department of Physics and Astronomy, LaserLaB, VU University Amsterdam,
De Boelelaan 1081, 1081 HV Amsterdam, The Netherlands}

\author{Wim Ubachs}%
\email[]{w.m.g.ubachs@vu.nl}
\affiliation{Department of Physics and Astronomy, LaserLaB, VU University Amsterdam,
De Boelelaan 1081, 1081 HV Amsterdam, The Netherlands}

\date{\today}

\begin{abstract}
Spontaneous Rayleigh-Brillouin scattering experiments in air, N$_2$ and O$_2$ have been performed for a wide range of temperatures and pressures at a wavelength of 403 nm and at a 90 degrees scattering angle. Measurements of the Rayleigh-Brillouin spectral scattering profile were conducted at high signal-to-noise ratio for all three species, yielding high-quality spectra unambiguously showing the small differences between scattering in air, and its constituents N$_2$ and O$_2$. Comparison of the experimental spectra with calculations using the Tenti S6 model, developed in 1970s based on linearized kinetic equations for molecular gases, demonstrates that this model is valid to high accuracy for N$_2$ and O$_2$, as well as for air. After previous measurements performed at 366 nm, the Tenti S6 model is here verified for a second wavelength of 403 nm, and for the pressure-temperature parameter space covered in the present study (250 --  340 K and 0.6 -- 3 bar).
In the application of the Tenti S6 model, based on the transport coefficients of the gases, such as thermal conductivity $\kappa$, internal specific heat capacity $c_{int}$ and shear viscosity $\eta$ as well as their temperature dependencies taken as inputs, values for the more elusive bulk viscosity $\eta_b$ for the gases are derived by optimizing the model to the measurements. It is verified that the bulk viscosity parameters obtained from previous experiments at 366 nm, are valid for wavelengths of 403 nm. Also for air, which is treated as a single-component gas with effective gas transport coefficients, the Tenti S6 treatment is validated for 403 nm as for the previously used wavelength of 366 nm, yielding an accurate model description of the scattering profiles for a range of temperatures and pressures, including those of relevance for atmospheric studies.
It is concluded that the Tenti S6 model, further verified in the present study, is applicable to LIDAR applications for exploring the wind velocity and the temperature profile distributions of the Earth's atmosphere. Based on the present findings at $90^{\circ}$ scattering and the determination of $\eta_b$ values predictions can be made on the spectral profiles for a typical LIDAR backscatter geometry.
These Tenti S6 predictions for Rayleigh-Brillouin scattering deviate by some 7\% from purely Gaussian profiles at realistic sub-atmospheric pressures occurring at 3-5 km altitude in the Earth's atmosphere.

\end{abstract}

%\keywords{}

\maketitle

\section{Introduction}

Knowledge of light scattering in gases dates back to the 19th century, when Lord Rayleigh (John William Strutt) explained that the blue sky was due to the scattering of sun light by gas molecules with diameters much less than the wavelength of the light~\cite{Strutt1899}. Using Maxwell's formalism of electromagnetism, Rayleigh successfully derived the scattering cross section as a function of the index of refraction of the gas and exhibiting the characteristic $\lambda^{-4}$ behavior.
%Driven by the electromagnetic field of the incident light, the electrons in the atoms or molecules radiate a secondary light with its phase exactly the same as the original one, thus this type of scattering by one molecule is coherent. Therefore, if all the molecules are homogeneously distributed without any movement, the coherent sum of the incident light and the scattered light cancels the radiation in all the other directions, and only the forward one remains. In real gases, however, the motion of the molecules leads to microscopic fluctuations and randomizes the phases of the scattered light. Hence, the scattering becomes incoherent.
Under the assumption of no collisions, molecular velocities follow a Maxwellian distribution and the spectral scattering profile exhibits a Gaussian line shape, produced from Doppler shifts associated with the molecular velocity distribution. In denser gases collisions occur between molecules, and Brillouin doublet peaks, Stokes and anti-Stokes shifted by the frequency of the acoustic waves, are formed and mix with the pure Rayleigh peak producing a complex Rayleigh-Brillouin (RB) scattering profile. According to Yip and Nelkin's work~\cite{Yip1964}, this spectral profile corresponds to the space-time Fourier-transform of the density-density correlation function $G(\boldsymbol{r},t)$. For compressed gases in the hydrodynamic regime, where many-body collisions frequently happen, $G(\boldsymbol{r},t)$ is represented as an ensemble average of density correlations~\cite{Mountain1966}. For diluted gases in the kinetic regime, where mainly two-body collisions occur, $G(\boldsymbol{r},t)$ is connected to the phase-space distribution function $f(\boldsymbol{r}, \boldsymbol{v}, t)$ in the Boltzmann equation~\cite{VanLeeuwen1965}. Therefore, the calculation of the spectral scattering profile of gases in the kinetic regime requires solving the Boltzmann equation. In view of  the mathematical difficulty in computing the collision integrals in the Boltzmann equation, kinetic models were developed to linearize it by assuming that only small deviations from equilibrium in the gaseous medium pertain~\cite{Boley1972,Tenti1974}. The Tenti S6 model~\cite{Tenti1974}, describing the collision integrals in 6 basis functions with their coefficients represented by the values of the macroscopic transport coefficients, such as thermal conductivity $\kappa$, shear viscosity $\eta$, bulk viscosity $\eta_b$, as well as the internal specific heat capacity per molecule $c_{int}$, has proven to be the most accurate model to represent the RB-scattering profile in the kinetic regime~\cite{Hubert1975,GhaemMaghami1980}. Furthermore, although this model is developed for gases in the kinetic regime, it is proven to be accurate in the hydrodynamic regime as well~\cite{Hammond1976}. % Also say something about the atomic gases? Not necessary here (WU).
Since the Tenti S6 model is mathematically involved, simpler analytical models consisting of 3 Gaussian~\cite{Witschas2011model} or 3 pseudo-Voigt functions~\cite{Ma2014} have been proposed, aiming to provide fast and simple representations of RB scattering spectra for remote sensing applications of the atmosphere. In any case the models must be validated against experiment, which is the purpose of the present study.

Experiments on RB spectral scattering profiles started in the 1960s, immediately after the invention of the laser as a source of narrow bandwidth radiation.
Brillouin doublet peaks, frequency-shifted from the central elastic Rayleigh peak, were detected both in liquids and solids~\cite{Benedek1964,Chiao1964}, as well as in gases~\cite{Greytak1966}. Subsequently, a number of studies were performed in molecular hydrogen~\cite{Hara1971}, molecular nitrogen~\cite{Sandoval1976}, various polyatomic gases~\cite{Lao1976}, and in noble gases~\cite{GhaemMaghami1980}, and compared with numerical models for the hydrodynamic regime~\cite{Mountain1966} %check
and for the kinetic regime~\cite{Boley1972,Tenti1974}. The latter studies led to the conclusion that the Tenti S6 model is the most successful approach in describing RB-scattering over a wide range of conditions~\cite{Young1983}.
Later, the research on RB scattering attracted less interest, until this century, when a new research technique known as coherent Rayleigh-Brillouin scattering (CRBS) was developed~\cite{Grinstead2000,Pan2002}.
Unlike the classical or spontaneous RB-approach, where the gas density fluctuations are spontaneously generated due to random thermal motion,
this coherent method uses two laser pulses to drive the fluctuations. After adding a term in the Boltzmann equation for the induced optical dipole force, kinetic models such as the S6 model can be used for describing the coherent RB scattering profiles~\cite{Pan2002,Pan2004}. The CRBS techniques were recently extended for measuring the temperature of a flame~\cite{Bookey2006flame} and for monitoring $\sim 1$ nm nano-particles in bulk gases or weakly ionized plasmas~\cite{Shneider2013}. %

Renewed research on spontaneous RB scattering is driven by the possible applications of LIDAR (light detection and ranging) techniques to obtain the wind speed distributions in the Earth's atmosphere, such as will be pursued by the ADM-Aeolus mission of the European Space Agency (ESA)~\cite{Stoffelen2006,Ansmann2007,Reitebuch2009}, for the temperature LIDAR experiments currently performed by German Aerospace Center (DLR)~\cite{Witschas2014Lidar}, as well as for aerosol LIDAR~\cite{Esselborn2008}.  Experimental measurements on RB scattering profiles of molecular gases, particularly of air as a gas mixture, are required to the highest possible signal-to-noise ratio to test the accuracy of the S6 model, which is proposed to be used in retrieval algorithms of LIDAR applications. Previous studies have proven that the Tenti S6 model is accurate to the 2\% level for a number of molecules, temperatures and pressures, at a wavelength of 366 nm, and for values of the bulk viscosity $\eta_{b}$ to be used in the S6 model as derived from RB-scattering experiments~\cite{Vieitez2010,ZiyuGu2013N2,ZiyuGu2013Air}.

The bulk viscosity is in principle a frequency or wavelength-dependent parameter, as has become evident from strongly varying values obtained from acoustic and optical measurements~\cite{Pan2005,ZiyuGu2014CO2}.
While the main purpose of the present study is to collect RB-scattering data for N$_2$, O$_2$ and air for atmospheric pressure and temperature conditions, values for the bulk viscosity will be derived for the additional wavelength 403 nm, where 366 nm was used in a previous study measuring RB-scattering in air~\cite{ZiyuGu2013Air} and in N$_2$~\cite{ZiyuGu2013N2}.
This is the reason why measurements are performed at elevated pressures up to 3 bar, under which conditions the Brillouin side peaks become more pronounced in the spectral profiles and a reasonably accurate determination of the bulk viscosity parameter $\eta_b$ is feasible. These values are of crucial importance for future modeling of RB-profiles under conditions of atmospheric LIDARS.
The present accurate measurements of RB-scattering in N$_2$, O$_2$ and air allow for a detailed comparison of the spectral profiles addressing the question whether air may be treated as a mono-molecular species as is done in the Tenti-framework.

\section{Theoretical models}

Light scattering is a result of fluctuations in a medium through which it propagates: in a completely homogeneous medium only forward scattering exists. The RB scattering phenomenon can be described by the elements of the dielectric tensor, representing fluctuations in thermodynamic quantities~\cite{Boyd2008}:
\begin{equation}
\label{equ:density_temperature}
 \Delta \epsilon = \Big( \frac{\partial \epsilon}{\partial \rho}\Big)_T \Delta \rho +
                    \Big( \frac{\partial \epsilon}{\partial T}\Big)_{\rho} \Delta T ,
\end{equation}
with the first term being the density fluctuations at constant temperature $T$ and second term the temperature fluctuations at constant density $\rho$.

Since scattering due to the temperature fluctuations of gases, corresponding to the second term in Eq.~(\ref{equ:density_temperature}), contributes only for $\sim2$\%~\cite{Fabelinskii1968}, this term is usually ignored. Furthermore, the entropy $s$ and pressure $p$ may be chosen to be independent thermodynamic variables representing the density fluctuations, thus yielding:
\begin{equation}
\label{equ:entropy_pressure}
 \Delta \rho = \Big( \frac{\partial \rho}{\partial p}\Big)_s \Delta p +
                    \Big( \frac{\partial \rho}{\partial s}\Big)_p \Delta s .
\end{equation}
The first term of Eq.~(\ref{equ:entropy_pressure}), describing pressure fluctuations also known as acoustic waves, results in  Brillouin scattering, while the second term describing entropy fluctuations, causes the Rayleigh scattering~\cite{Boyd2008}.
For gases in the kinetic regime, where two-body collisions dominate, the Boltzmann equation is adequate to describe the density fluctuations. Since the collision integral of the Boltzmann equation is difficult to compute, models of RB scattering in molecular gases based on the linearized Wang-Chang-Uhlenbeck equation~\cite{Wang-Chang1970},
a modified version of the Boltzmann equation for molecular gases, have been developed.
In these models the Boltzmann equation is cast into seven~\cite{Boley1972} or six~\cite{Tenti1974} matrix elements (now regarded as the S7 or S6 model, respectively), which are directly related to the macroscopic transport coefficients, namely the shear viscosity $\eta$, the thermal conductivity $\kappa$, the bulk viscosity $\eta_b$, as well as the internal specific heat capacity per molecule $c_{int}$.
Therefore, based on experimental knowledge of these coefficients at specific temperatures and pressures, and inserting the molecular mass of the gas constituents, RB scattering profiles can be calculated by the S6 or S7 models.

In this framework air has been successfully treated as a single-component gas with an effective particle mass 29.0 u and with effective transport coefficients obtained from independent measurements~\cite{Rossing2007}. For diatomic gases, such as N$_2$, O$_2$, or air with its major components being N$_2$ and O$_2$, $c_{int}$ is always equal to 1. The shear viscosity $\eta$ and thermal conductivity $\kappa$ are known to be independent of pressure~\cite{White1998} (especially for the pressure ranges employed in the present study). However, these transport coefficients are dependent on temperature, following the Sutherland formulas~\cite{White1991}:
\begin{equation}\label{equ:shear_viscosity}
\frac{\eta}{\eta_0} = \left( \frac{T}{T_0} \right)^{3/2}  \frac{T_0+S_{\eta}}{T+S_{\eta}},
\end{equation}
and
\begin{equation}\label{equ:thermal_conductivity}
\frac{\kappa}{\kappa_0} = \left( \frac{T}{T_0} \right)^{3/2}  \frac{T_0+S_{\kappa}}{T+S_{\kappa}},
\end{equation}
where $\eta_0$ is the reference shear viscosity and $\kappa_0$ the reference thermal conductivity, at reference temperature $T_0$ (normally 273 K) and $S_{\eta}$ and $S_{\kappa}$ are called Sutherland parameters. Values for $\eta_0$, $\kappa_0$, $T_0$, $S_{\eta}$ and $S_{\kappa}$ are adopted from~\cite{White1991} for N$_2$, O$_2$, and air, and listed in Table~\ref{Const:Sutherland}. All Tenti S6 calculations in the present study are based on the transport coefficients derived from these constants, as well as on a value $c_{int}=1$ for the internal specific heat capacity.

\begin{table}
{\caption{\label{Const:Sutherland} Coefficients for the calculation of the shear viscosity $\eta$ and the thermal conductivity $\kappa$ of air, N$_2$ and O$_2$ via the Sutherland formulas Eqs.~(\ref{equ:shear_viscosity}) and (\ref{equ:thermal_conductivity}).
}}
\begin{center}
\begin{tabular}{c c c c c c}
\hline
                    &  $T_0$    &   $\eta_0$   &  $\kappa_0$          & $S_{\eta}$    & $S_{\kappa}$   \\
                    & (K)       &    (kgm$^{-1}$s$^{-1}$)   & (WK$^{-1}$m$^{-1}$)   & (K) &  (K) \\
\hline
     Air            &  273     &   $1.716 \times 10^{-5}$    &  0.0241    &   111   &   194 \\%20130810
    N$_2$           &  273     &   $1.663 \times 10^{-5}$    &  0.0242    &   107   &   150 \\%20130804
     O$_2$          &  273     &   $1.919 \times 10^{-5}$    &  0.0244    &   139   &   240  \\%20130803
\hline
\end{tabular}
\end{center}
\end{table}

Another transport coefficient required for the S6 modeling, the bulk viscosity $\eta_b$, is related to the energy exchanges between the translational and internal (rotational and vibrational) degrees of freedom of molecules through collisions. For mono-atomic gases, it is straightforward to set $\eta_b=0$, since there are no internal degrees of freedom~\cite{Tisza1942}. For molecular gases, on the other hand, it must be considered how many internal degrees of freedom effectively contribute to the bulk viscosity. As pointed out by Meijer~\emph{et al.}~\cite{Meijer2010}, the bulk viscosity is determined by the product of $\omega \tau_j$, where $\omega$ is the angular frequency of sound waves and $\tau_j$ is the relaxation time of the internal mode $j$. In the extreme case of $\omega \tau_j \to  \infty$, the fluctuations resulting in sound waves are so fast that there is no energy transfer between the sound-driven translational motion and the motion of internal mode $j$, so mode $j$ is frozen and there is a zero contribution to the bulk viscosity. In contrast, if $\omega \tau_j << 1$, there are sufficient collisions within one wave period to maintain local thermodynamic equilibrium, and the contribution of mode $j$ to the bulk viscosity is frequency-independent. In general cases, however, the bulk viscosity is known as a frequency-dependent parameter. Of course, $\eta_b$ is a temperature-dependent parameter, since at higher temperatures more degrees of freedom will participate in the internal motion of the molecules and the relaxation time for the internal motion are shorter as collisions more frequently happen~\cite{Cramer2012}.

Most information about the numerical values of the bulk viscosity $\eta_b$ for diluted gases comes from ultrasound experiments at MHz frequencies~\cite{Prangsma1973}. Since the sound frequency in light-scattering experiment is  $|\boldsymbol{k_s}-\boldsymbol{k_i}|v \sim 1$ GHz (with $\boldsymbol{k_i}$ and $\boldsymbol{k_s}$ being the wave vector of the incident and scattered light, and $v$ the speed of sound in the gases), it is questionable whether the values measured at low frequencies can be directly used in the S6 model. Indeed, Pan \emph{et al.}~\cite{Pan2004,Pan2005} found a 3 orders of magnitude discrepancy for $\eta_b$ values derived from acoustic (ultrasound) and optical (coherent RBS) experiments for the case of CO$_2$. For these studies Pan \emph{et al.} had proposed that values of the bulk viscosity at hypersonic frequencies can be measured by RB-scattering experiments through comparison between the measured and calculated scattering profiles, for the reason that the bulk viscosity is the only uncertain parameter in the Tenti S6 model. Measurements on spontaneous RB-scattering, also in comparison with the Tenti S6 model, validated the hypersonic $\eta_b$ value for CO$_2$~\cite{ZiyuGu2014CO2}. This large discrepancy between $\eta_b$ values at ultrasound (MHz) and hypersonic (GHz) frequencies is attributed to the relatively slow relaxation time for vibrational motion of CO$_2$ at atmospheric pressures, which is $\tau_v = 6\times 10^{-6}\; {\rm s}$. For light scattering experiments, probing the hypersound domain,  $\omega \tau_v \approx 1000$, so the vibrational modes are frozen and the bulk viscosity is much smaller than the values obtained from sound absorption measurements with $\omega \tau_v \approx 1$.  Meijer~\emph{et al.}~\cite{Meijer2010} recorded values of $\eta_b$ for various gases at room temperatures using coherent RB scattering at 532 nm, and compared them with the values from acoustic measurements and molecular structure calculations, indicating that for polyatomic gases the values at hypersonic frequencies are generally smaller than at ultrasonic frequencies. Gu~\emph{et al.} extended the study to a range of temperatures~\cite{ZiyuGu2013Air,ZiyuGu2013N2} using spontaneous RB scattering at 366 nm, demonstrating that the bulk viscosity increases with the temperature, but is insensitive to a change of pressure, at least in the regime up to 3 bar. In the present work values for the bulk viscosity for air, N$_2$, and O$_2$ are derived from spontaneous RB-scattering experiments at 403 nm for a range of temperatures.

\section{Experimental}
\begin{figure}[ht]
\includegraphics[width=8cm]{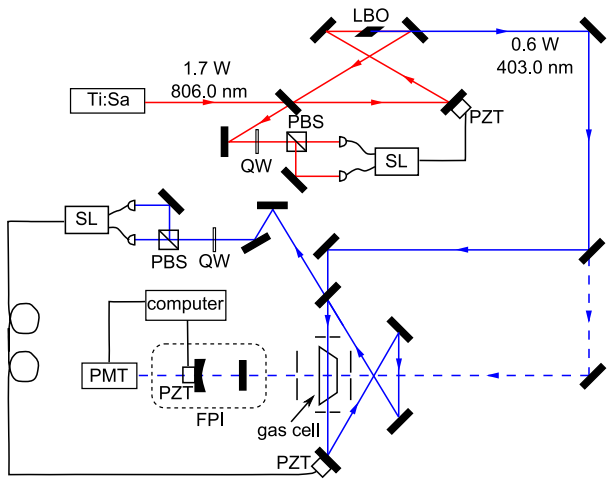}
\caption{\label{fig:403_Setup} Layout of the experimental apparatus. The Ti:Sa laser beam, pumped by a 10 W 532 nm Millennia Xs laser, is intracavity frequency-doubled in a Lithium-Borate (LiB$_3$O$_5$ or LBO) crystal, yielding a cw power of 600 mW at 403 nm. The blue laser beam is then directed into a second enhancement cavity for amplification by a factor of 10, in which the RB-scattering cell is placed to ensure a maximum scattering intensity. The polarizing beam splitters (PBS), quarter wave plates (QW), piezo tubes (PZT) and servo loops (SL) are required elements to lock these two cavities. Scattered light is collected at $90^{\circ}$ with respect to the beam direction. The geometrically filtered scattered light is directed onto a Fabry-Perot Interferometer (FPI), where transmitted photons are detected by a photo-multiplier tube (PMT). A small fraction of the 403 nm light is used as a reference beam for aligning beam paths and for characterizing the detecting system (dashed lines). This reference beam is blocked when measurements are performed. }
\end{figure}

Fig.~\ref{fig:403_Setup} displays a sketch of the experimental setup used for this study, exhibiting the same geometry as the one reported in~\cite{ZiyuGu2012} for RB scattering experiments at 366 nm. Some relevant details on the experimental parameters are specified in the caption. Due to the significant change of the wavelength (from 366 nm to 403 nm), all optical components are replaced, and the alignment is readjusted.

RB-scattering measurements are performed for various pressures and temperatures in a $p-T$ parameter space following the experimental procedure outlined here.
Before each measurement, the scattering cell is charged with one of the sample gases (air, N$_2$ or O$_2$) to one of the three approximate initial charging pressures, 1000 mbar, 2000 mbar, and 3000 mbar, at room temperature. After sealing the cell, the temperature is set to a designated value, with 0.5 K uncertainty. Hence the pressure of the gas inside the scattering cell is changed and its value is calculated from the ideal gas law.  The cell allows for pressure settings of the sample gas between 0 -- 4 bar with 0.5\% uncertainty.
The temperature of the sample gas inside the scattering cell can be controlled and stabilized within the range 250 K -- 350 K. The actual $p-T$ settings for the measurements are listed in Table~\ref{Tab:conditions}.
Throughout the paper the wavelength will be referred to as 403 nm, while the exact wavelength is always set in the window $\lambda = 402.99 - 403.00$ nm, as measured with a wavelength meter (ATOS).

\begin{table*}[th]
{\caption{\label{Tab:conditions} Conditions and values of transport coefficients for the Rayleigh-Brillouin scattering measurements for air, N$_2$ and O$_2$ at pressures $p$ and temperatures $T$ as indicated. Values for $\eta$ and $\kappa$ are calculated by Eq.~(\ref{equ:shear_viscosity}) and Eq.~(\ref{equ:thermal_conductivity}), using the constants given in Table~\ref{Const:Sutherland}. Values for the bulk viscosity $\eta_b$ are obtained from the data at $\sim 3$ bar by fitting to the Tenti S6 model, while for other pressure-temperature conditions values are derived by the interpolation procedure discussed in section~\ref{sec:bulk}.  Also specified is the angle $\theta$ at which the data are recorded. All data measured at $\lambda$ in the range 402.99 - 403.00 nm. Data files for calibrated data, indicated by file name in the last column, are made publicly available via the Supplementary Material~\cite{Suppl}.
}}
\begin{center}
\begin{tabular}{c c c c c c c l}
\hline
      & $p$        & $T$         & $\eta$    & $\eta_b$ & $\kappa$ & $\theta$   & Datafile\\
      & (mbar)   & (K)       & (10$^{-5}$ kgm$^{-1}$s$^{-1}$) & (10$^{-5}$ kgm$^{-1}$s$^{-1}$) & (10$^{-2}$ WK$^{-1}$m$^{-1}$) & degrees \\
\hline
\hline
                    &  880      &   256.6   &   1.633   &   0.909    &   2.279    & 91 & A1257K \\ %
                    &  953      &   278.3   &   1.741   &   1.252    &   2.453    & 91 & A1278K \\ %
 Air $\sim1$ bar    &  1012     &   295.8   &   1.826   &   1.527    &   2.591    & 91.5 & A1296K \\ %
                    &  1095     &   320.1   &   1.940   &   1.913    &   2.779    & 91.5 & A1320K \\ %
                    &  1165     &   339.9   &   2.029   &   2.226    &   2.928    & 91.5 & A1320K \\ %
 \hline
                    &  1898     &   280.8   &   1.754   &   1.291   &   2.474     & 90.5 & A2281K \\ %
                    &  2000     &   295.5   &   1.825   &   1.524   &   2.589     & 90.5 & A2296K \\ %
Air $\sim2$ bar     &  2131     &   314.9   &   1.916   &   1.831   &   2.739     & 90.5 & A2315K \\ %
                    &  2300     &   339.9   &   2.029   &   2.226   &   2.928     & 90.5  & A2340K \\ %
\hline
                    &  2604    &   254.5   &   1.622   &   0.905   &   2.263    & 89.5 & A3255K \\%
                    &  2831    &   279.6   &   1.748   &   1.220   &   2.464    & 90.5 & A3280K \\%
Air $\sim3$ bar     &  3000    &   296.8   &   1.831   &   1.580   &   2.599    & 90.5 & A3297K \\%
                    &  3196    &   315.7   &   1.919   &   1.805   &   2.745    & 90.5 & A3316K \\%
                    &  3444    &   340.2   &   2.030   &   2.255   &   2.930    & 90.5 & A3340K \\%
\hline
\hline
                    &  863     &   253.8   &   1.570   &   0.751   &   2.272   & 91 & N1254K  \\%
                    &  945     &   277.0   &   1.682   &   1.121   &   2.441   & 91 & N1277K   \\%
N$_2$ $\sim1$ bar   &  948     &   296.3   &   1.772   &   1.428   &   2.593   & 91.5 & N1296K  \\%
                    &  1010     &   315.7   &   1.859   &   1.738   &   2.733  & 91.5 & N1316K  \\%
                    &  1093     &   341.2   &   1.970   &   2.144   &   2.912  & 91.5 & N1341K  \\%
 \hline
                    &  1898     &   280.8   &   1.700   &   1.181   &   2.479  &   90.5    & N2281K   \\%
                    &  2000     &   295.8   &   1.774   &   1.420   &   2.597  &   90.5  & N2296K   \\%
N$_2$ $\sim2$ bar   &  2133    &   315.7   &   1.859   &   1.738    &   2.733   &  90.5 & N2316K   \\%
                    &  2275    &   336.5   &   1.950   &   2.069   &   2.880   &  90.5 & N2337K  \\%
\hline
                    &  2589    &   254.8   &   1.575   &   0.793   &   2.279  &  90 & N3255K \\%
                    &  2828    &   279.8   &   1.695   &   1.130   &   2.471  &  90 & N3280K \\%
N$_2$ $\sim3$ bar   &  3000    &   297.3   &   1.776   &   1.400   &   2.601  &  90 & N3297K  \\ %
                    &  3194    &   315.4   &   1.858   &   1.805   &   2.732  &  90 & N3315K \\ %
                    &  3417    &   338.1   &   1.957   &   2.075   &   2.890  &  90.5 & N3338K  \\%
\hline
\hline
                    &  1000     &   295.5   &   2.033   &   1.291   &   2.591  & 91.5  & O1296K  \\%
                    &  1150     &   339.7   &   2.180   &   2.633   &   2.778  & 91.5 & O1340K  \\%
O$_2$               &  2000    &   295.8   &   2.034   &   1.300   &   2.593   & 90   & O2296K \\%
                    &  2270    &   335.7   &   2.167   &   2.512   &   2.762   & 90   & O2336K  \\ %
                    &  3000    &   297.6   &   2.040   &   1.355   &   2.600   & 90   & O3298K \\%
                    &  3419    &   339.1   &   2.178   &   2.615   &   2.776   & 90   & O3339K \\%
\hline
\hline
\end{tabular}
\end{center}
\end{table*}

While Rayleigh-Brillouin scattering measurements were intended for a right angles geometry, the actual scattering angles depended on the final adjustment of the alignment of laser-beam and the beam-path of the scattered light. All the initial measurements for 1 bar (in air, N$_2$, and O$_2$) were performed for a scattering angle of 91.5$^{\circ} \pm 0.9^{\circ}$, while for the later measurements at other pressures the scattering was re-adjusted to 90$^{\circ} \pm 0.9^{\circ}$.

\begin{figure}
\includegraphics[width=8cm]{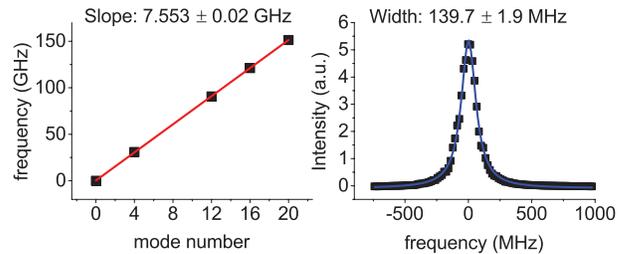}
\caption{\label{fig:FPI} Left: Measurement series of frequencies of transmission fringes of the Fabry-Perot interferometer while scanning the laser and calibration by a wavelength meter; from the span of 20 modes the FSR is determined. Right: Recording of a single transmission fringe of the FPI for a measurement of the instrument profile; while this single measurement yields $139\pm1.9$ MHz, an average over multiple measurements delivers the instrument width of $140\pm4$ MHz to be used in the analyses.}
\end{figure}

The scattered photons are spectrally resolved by a home-built plano-concave Fabry-Perot interferometer (FPI) that was characterized by scanning the laser over an extended range covering a number of free-spectra-ranges (FSR), while monitoring the wavelength on the wavelength meter (ATOS). This procedure, further detailed in~\cite{ZiyuGu2012}, delivers an accurate effective value for the FSR of 7553 MHz, when covering a span of 20 modes (see Fig.~\ref{fig:FPI}). The FSR-scale was used to calibrate the laser scan in a recording of the profile of the spectral profile of an individual transmission fringe, yielding an instrument linewidth of $140\pm4$ MHz, when averaging over multiple calibrations.  A high gain photo-multiplier tube (PMT) is used to detect the photons passing through the FPI. The RB-spectral profiles are recorded by keeping the laser frequency fixed, while scanning the FPI with a piezo for a typical period of 3 hours. During such observation a frequency span of 7500 GHz (corresponding to $\sim100$ FSRs) is covered. Measures for correcting the drift of the laser frequency and the FPI are applied, followed by linearizing, averaging and normalizing the $\sim100$ resolved scattering line shapes to area unity~\cite{ZiyuGu2012}.

Finally, the normalized scattering profiles are compared with the numerical S6 calculations, performed for the exact measurement conditions, and convolved with the instrument function of the FPI, which is written as:
\begin{equation}
A=I_0 \cdot \frac{1}{1+(\frac{2\cdot FSR}{\pi \cdot FWHM})^2 \cdot \textrm{sin}^2(\frac{\pi}{FSR} \cdot f)},
\end{equation}
where $f$ is the frequency, and values of $FSR=7553$ MHz and $FWHM=140$ MHz are included.

It is worth mentioning that previous measurements reported in~\cite{Vieitez2010,ZiyuGu2013N2,ZiyuGu2013Air} suffered from background problems, in that the side wings of the measured scattering profiles were found to be higher than the side wings of the Tenti S6 calculations, even though the dark counts of the PMT and the overlap of the scattering profiles between two FSRs were accounted for. This background was firstly ascribed to broadband fluorescence of the cell windows in~\cite{Vieitez2010}, and was later understood as an influence of Raman scattering~\cite{ZiyuGu2013Air}. In order to investigate the influence of Raman scattering, most of which maintains a large frequency shift from the incident light, a high transmission (90\%) narrowband ($\Delta\lambda = 1$~nm) bandwidth filter with its central wavelength at 403 nm is implemented in front the PMT in this study to block most of the Raman scattering. Indeed, it is found that the measurements reported in the present study no longer suffer from Raman-associated background problems.

\begin{figure}
\includegraphics[width=8cm]{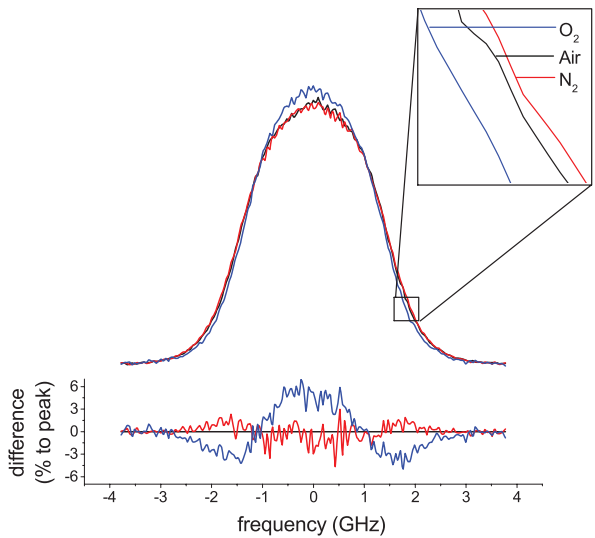}
\caption{\label{fig:Gascomparison} Measurement of the Rayleigh-Brillouin scattering spectral profiles for N$_2$, O$_2$ and air at 1 bar pressure, room temperature, $\lambda=403.00$ nm and $\theta=91.5^{\circ}$. In the bottom part differences between N$_2$-air (red) and O$_2$-air (blue) are plotted.}
\end{figure}

To illustrate the sensitivity of the Rayleigh-Brillouin spectrometer in Fig.~\ref{fig:Gascomparison} typical recordings of the RB-profile for the three gases, N$_2$, O$_2$ and air are shown for conditions of 1 bar, room temperature, $\lambda=403.00$ nm and a scattering angle of $\theta=91.5^{\circ}$. While these spectra, all three normalized to unity, correspond to top-lowered Gaussian-like profiles, the shapes are mainly determined by the masses of the constituent molecules through the Doppler effect. Where N$_2$ has a mass of 28 u, O$_2$ has 32 u, and air may be treated as a species of effective mass 29 u. Indeed the profile for O$_2$ is narrower and hence exhibits an increased intensity near the line centre, while the profile of N$_2$ is only slightly broader (and lower in the centre) than that of air. But the differences are still measurable, thus demonstrating the sensitivity of the RB-spectrometer.

%\subsection{Air}
\begin{figure}
\includegraphics[width=8.3cm]{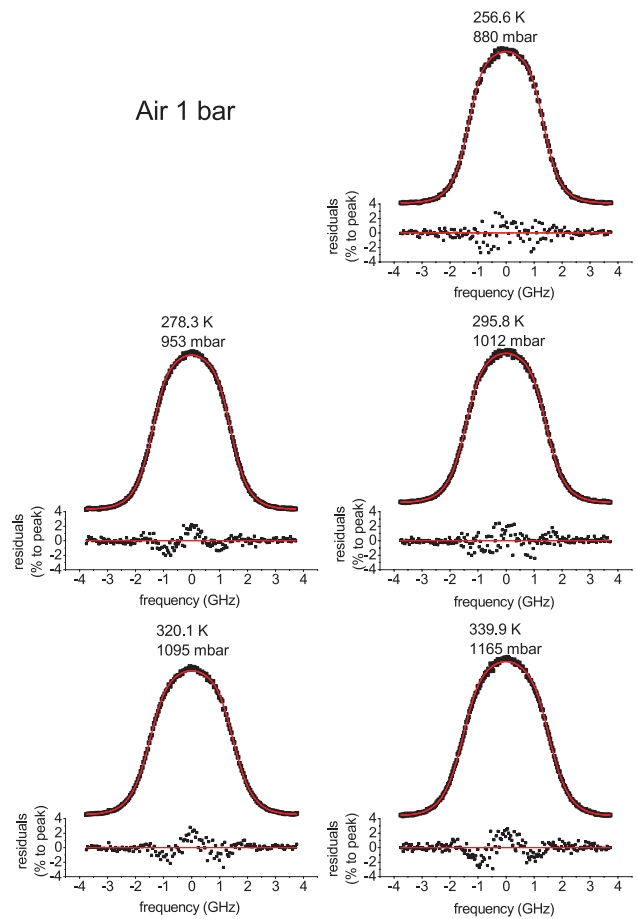}
\caption{\label{fig:air_1bar_403nm_all}
Normalized Rayleigh-Brillouin scattering profiles of air recorded at $\lambda=403.0$ nm (black dots), for pressures $\sim 1000$ mbar and temperatures as indicated. The scattering angle for this data was $\theta = 91.7^\circ$. Experimental data (black dots) are compared with the convolved Tenti S6 model calculations (red line), with values of $\eta$, $\kappa$, and $\eta_b$ as input parameters listed in Table~\ref{Tab:conditions}.}
\end{figure}

\begin{figure}
\includegraphics[width=8.3cm]{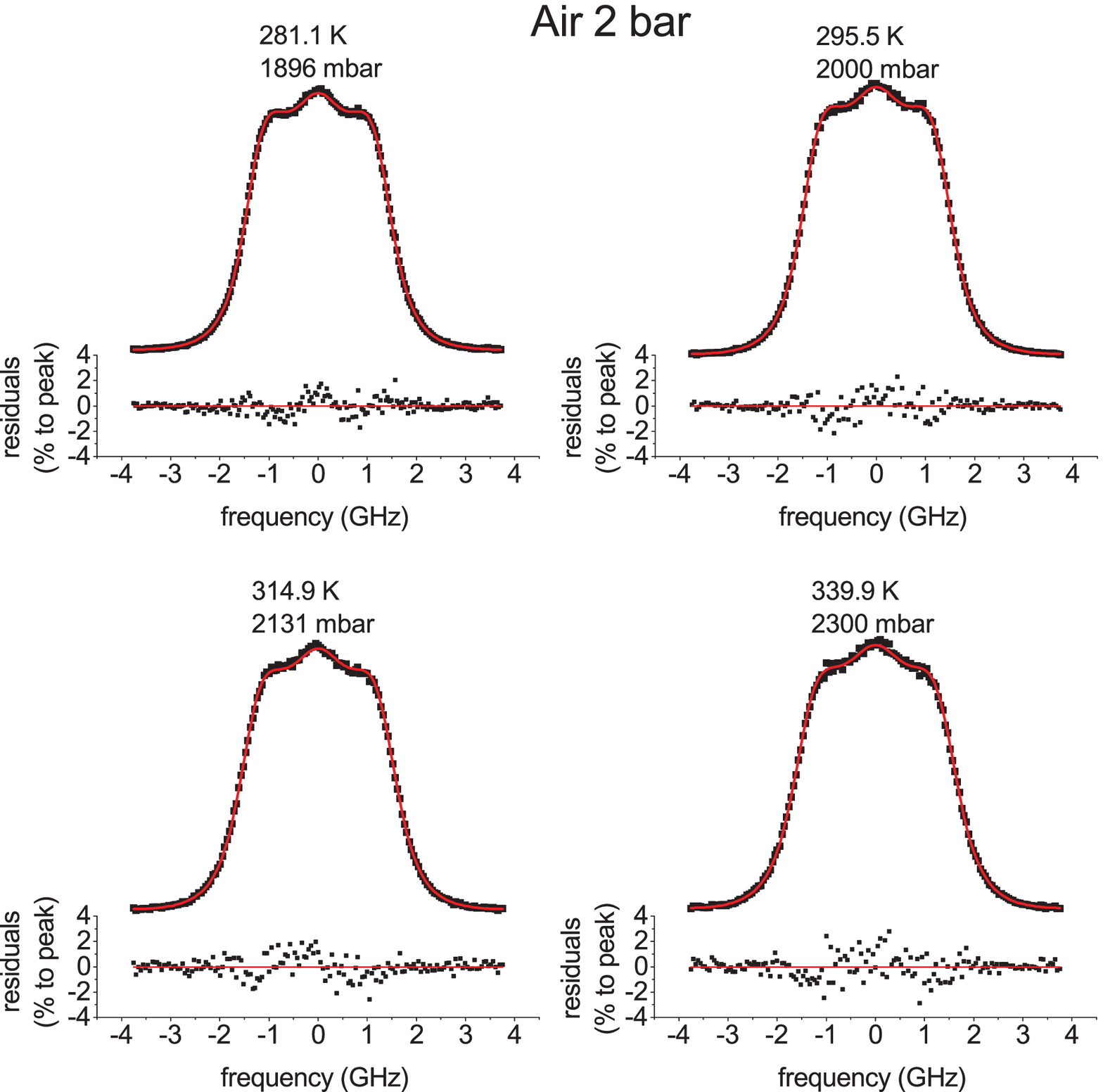}
\caption{\label{fig:air_2bar_403nm_all} Normalized Rayleigh-Brillouin scattering profiles of air (black dots) recorded for pressures $\sim 2000$ mbar and temperatures as indicated. The scattering angle for this data set was $\theta = 90.0^\circ \pm 0.9^\circ$.}
\end{figure}

\begin{figure}
\includegraphics[width=8.3cm]{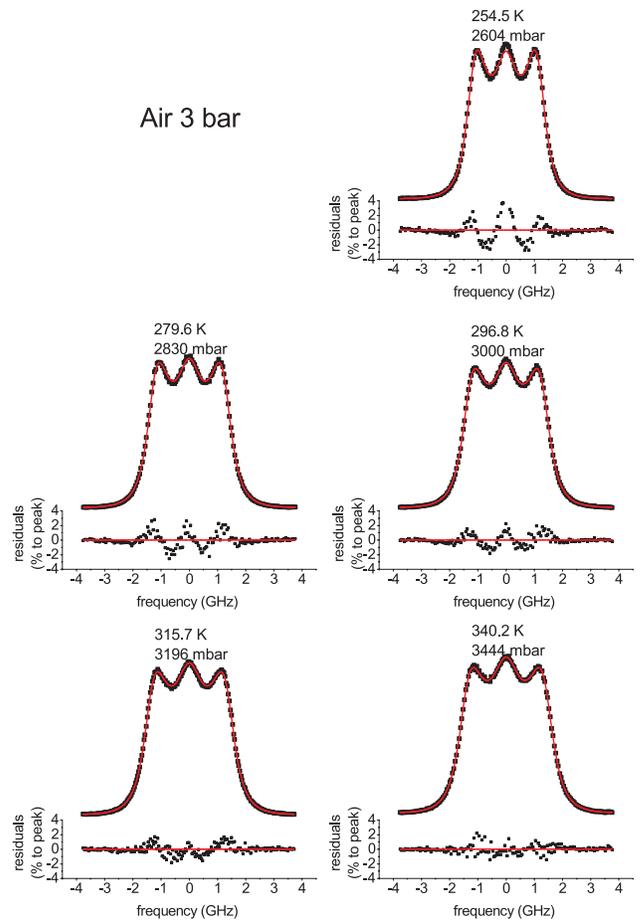}
\caption{\label{fig:air_3bar_403nm_all} Normalized Rayleigh-Brillouin scattering profiles of air (black dots) recorded for pressures $\sim 3000$ mbar. The scattering angle for all the measurements in this figure is $\theta = 90.0^\circ \pm 0.9^\circ$. Values of $\eta$, $\kappa$ are listed in Table~\ref{Tab:conditions}, while values of $\eta_b$ at different temperatures are directly obtained from the least-squared fit to the S6 model (red curves).}
\end{figure}

\section{Measurements and Analysis}

Comprehensive data sets on measurements of spontaneous Rayleigh-Brillouin scattering in air, N$_2$, and O$_2$ at different temperatures and pressures are reported. A choice was made to record spectra for three different initial charging pressures, 1000 mbar, 2000 mbar and 3000 mbar, combined with five different temperature settings at 255 K, 277 K, 297 K, 318 K and 337 K, at intervals of $\sim$20 K. The conditions under which the data were recorded are listed in Table~\ref{Tab:conditions}. The raw data, after linearization and calibration on a frequency scale, and subsequent averaging, are made available in Supplementary Material to this paper.

%\subsection{N$_2$}

\begin{figure}
\includegraphics[width=8.3cm]{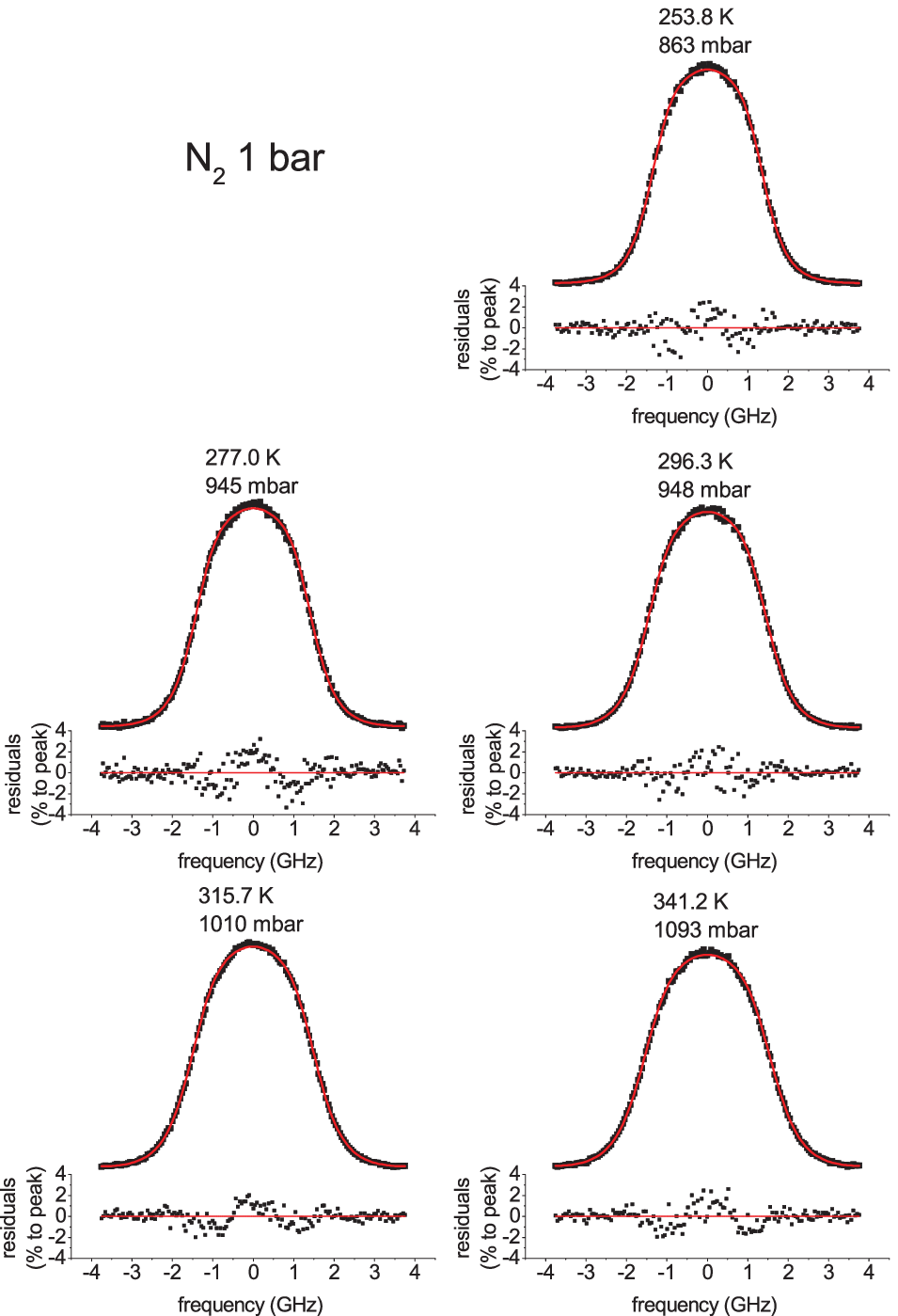}% rewrite later.
\caption{\label{fig:N2_1bar_403nm_all} Normalized Rayleigh-Brillouin scattering profiles of N$_2$ recorded at $\lambda=403.0$ nm, for pressures $\sim 1000$ mbar and temperatures as indicated. The scattering angle for this data was $\theta = 91.7^\circ$. Experimental data (black dots) are compared with the convolved Tenti S6 model calculations (red line), with the input parameters, $\eta$, $\kappa$, and $\eta_b$, listed in Table~\ref{Tab:conditions}.}
\end{figure}

\begin{figure}
\includegraphics[width=8.3cm]{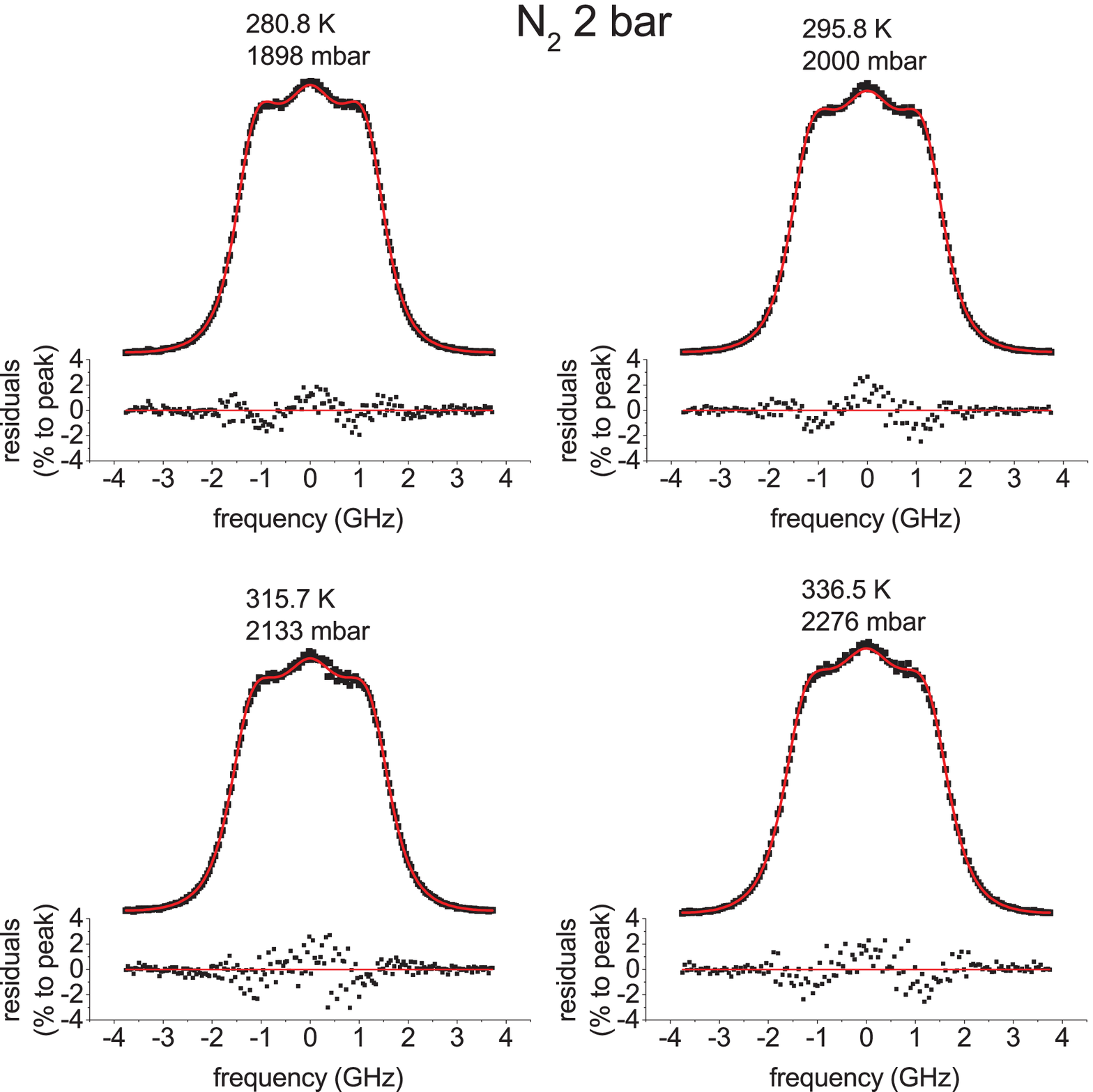}% rewrite later.
\caption{\label{fig:N2_2bar_403nm_all} Normalized Rayleigh-Brillouin scattering profiles of N$_2$ (black dots) recorded for pressures $\sim 2000$ mbar and temperatures as indicated. The scattering angle for this data set was $\theta = 90.0^\circ \pm 0.9^\circ$.}
\end{figure}

\begin{figure}
\includegraphics[width=8.3cm]{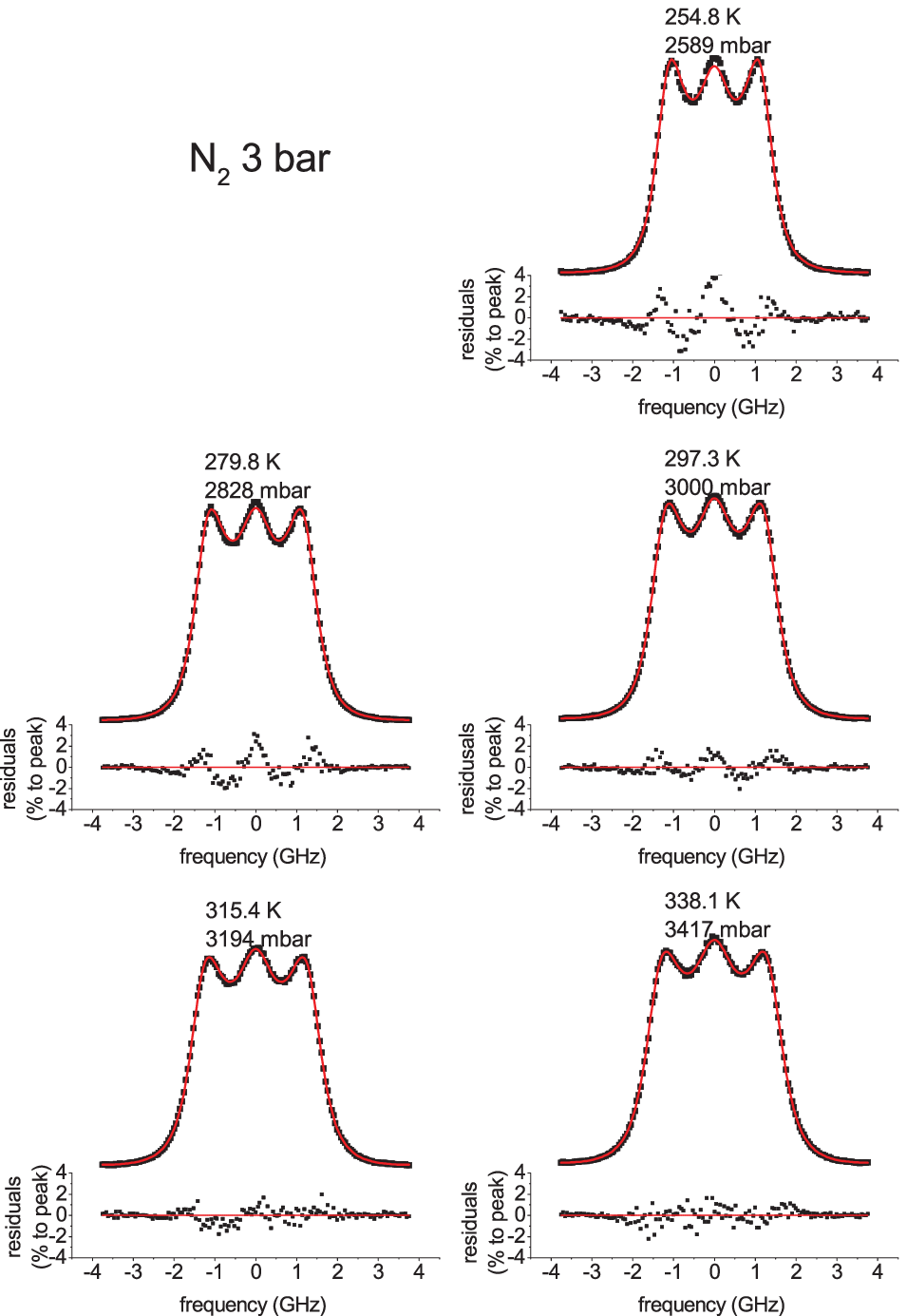}% rewrite later.
\caption{\label{fig:N2_3bar_403nm_all} Normalized Rayleigh-Brillouin scattering profiles of N$_2$ (black dots) recorded for pressures $\sim 3000$ mbar. The scattering angle for all the measurements in this figure was $\theta = 90.0^\circ \pm 0.9^\circ$. Values of $\eta$, $\kappa$ are listed in Table~\ref{Tab:conditions}, while values of $\eta_b$ at different temperatures are directly obtained from the least-squared fit of the S6 model (red curves) to the measurements.}
\end{figure}

%\subsection{O$_2$}

\begin{figure}
\includegraphics[width=8.3cm]{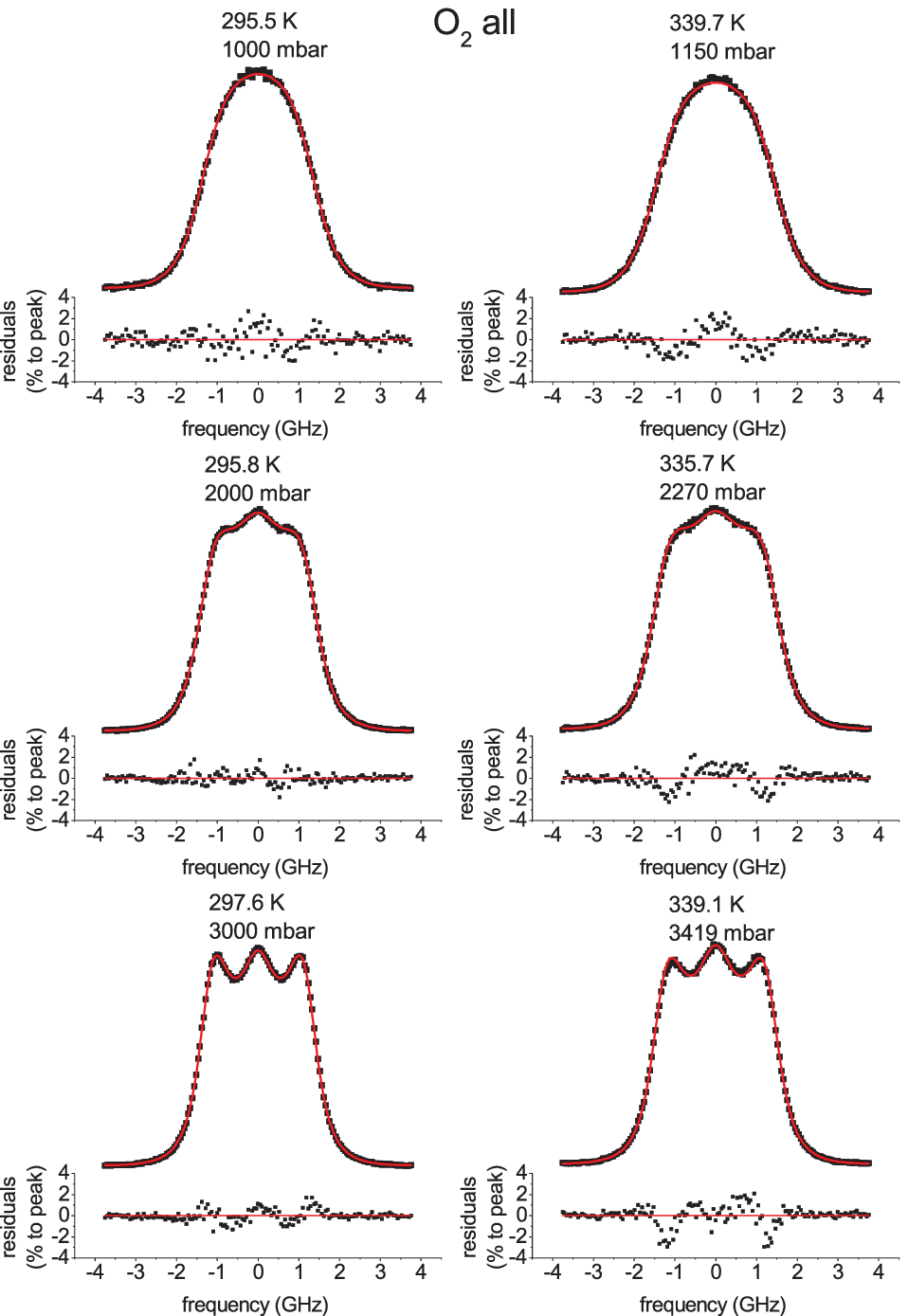}% rewrite later.
\caption{\label{fig:O2_403nm_all} Normalized Rayleigh-Brillouin scattering profiles of O$_2$ recorded for pressures and temperatures as indicated. For the two measurements on the first row, corresponding to pressures $\sim1$ bar, the scattering angle is  $\theta = 91.7^\circ$, while for the measurements in the lower two rows, the scattering angle is $90.0^\circ \pm 0.9^\circ$.}
\end{figure}

Spectral recordings for Rayleigh-Brillouin scattering under different $p-T$ conditions and for the three gases is shown in a series of figures, from Fig.~\ref{fig:air_1bar_403nm_all} to Fig.~\ref{fig:O2_403nm_all}, with results from calculations using the S6 model overlaid, and residuals (in percentage of the peak of the scattering profiles) plotted underneath each measurement. Values of the gas transport coefficients implemented in the S6 model, according to Eq.~(\ref{equ:shear_viscosity}) and Eq.~(\ref{equ:thermal_conductivity}), are listed in Table~\ref{Tab:conditions}. The bulk viscosity $\eta_b$ at hypersonic frequencies is treated as a unknown parameter, and is derived from fitting to the subset of experimental data obtained for pressures of 3 bar.
Since the bulk viscosity is associated with collisions it has a more pronounced effect at higher pressures.
Therefore, the $\eta_b$ values for air, N$_2$ and O$_2$ are obtained from spectra recorded at 3 bar, from a least-squares fitting procedure to the S6 model, using~\cite{Meijer2010}:
\begin{equation}
\label{equ:least-square fit}
\chi ^2=\frac{1}{N} \sum_{i=1}^N \frac{[I_e(f_i)-I_m(f_i)]^2}{\sigma^2(f_i)},
\end{equation}
where $I_e(f_i)$ and $I_m(f_i)$ are the experimental and modeled amplitude of the spectrum at frequency $f_i$, and $\sigma(f_i)$ the statistical (Poisson) error~\cite{ZiyuGu2013Air}.
Since it follows from previous studies, that the bulk viscosity exhibits a linearly increasing trend with the temperature~\cite{ZiyuGu2013Air,ZiyuGu2013N2}, as it is expected from theoretical considerations~\cite{Cramer2012},
values of the bulk viscosity are derived for the different temperature settings for the experiments, for the three different gases.
The thus obtained values for the derived bulk viscosities for air, N$_2$ and O$_2$ as a function of temperature (for further discussion see below) are employed to interpolate the values for the other settings in $p-T$ parameter space as listed in Table~\ref{Tab:conditions}. Here it is assumed that the bulk viscosities are pressure-independent, and that for S6 calculations at lower pressures (1 bar and 2 bar) the values for $\eta_b$ derived from measurements at 3 bar can be used.

For all the measurements shown in Fig.~\ref{fig:air_1bar_403nm_all} to Fig.~\ref{fig:O2_403nm_all}, the measurement noise is $\sim 1$\% of their peak intensity, while the difference between the measurements and the calculations is $\sim 2$\%, at approximately the same difference as reported in~\cite{ZiyuGu2013Air,ZiyuGu2013N2}.
In previous analyses of RB-scattering at 366 nm the elastic scattering from dust particles (Mie scattering) or scattering from the inner walls of the cell had been identified as sources of deviations at the center of the RB-spectrum~\cite{ZiyuGu2013Air}. Such effects should appear at widths of 140 MHz, corresponding to the instrument bandwidth, but in fact appear as much broader features in the present spectra. This reduces the possibility of elastic scattering contributing in the present measurements, where indeed special care was taken to avoid dust entering the cell.

Rotational Raman scattering, contributing to $\sim2.5$\% of the total cross section, is not considered in the Tenti S6 model, hence could be another possible source for the deviation. However, since a 1 nm bandwidth filter is used for all the measurements presented here, most of the rotational Raman scattering should be filtered out, and indeed the previously detected problems with baseline intensities no longer pertain for the data set recorded at 403 nm. There exists a special form of rotational Raman scattering, corresponding to no change of the rotational levels of the gas molecules (i.e. $\Delta J=0$) but a change of the projection of the rotational angular momentum of the molecules on a space-fixed axis (i.e. $\Delta m_J=\pm2$)~\cite{Miles2001}. Such scattering processes should reproduce the entire RB-profile, since no experimental distinction between Rayleigh-Brillouin and elastic Raman scattering can be made based on the spectral profile.
Such elastic Raman contributions, however, produce depolarized light, a phenomenon which might be subject of future investigations. Additionally, it is worth noticing that the temperature fluctuations of gases, possibly contributing to some 2\% of the scattering intensity, is neglected in making the step from Eq.~(\ref{equ:density_temperature}) to Eq.~(\ref{equ:entropy_pressure}). Also this effect should in first order affect the scattering intensity rather than the RB-spectra profile.

\section{Bulk viscosity}
\label{sec:bulk}

\begin{figure}
\includegraphics[width=8.3cm]{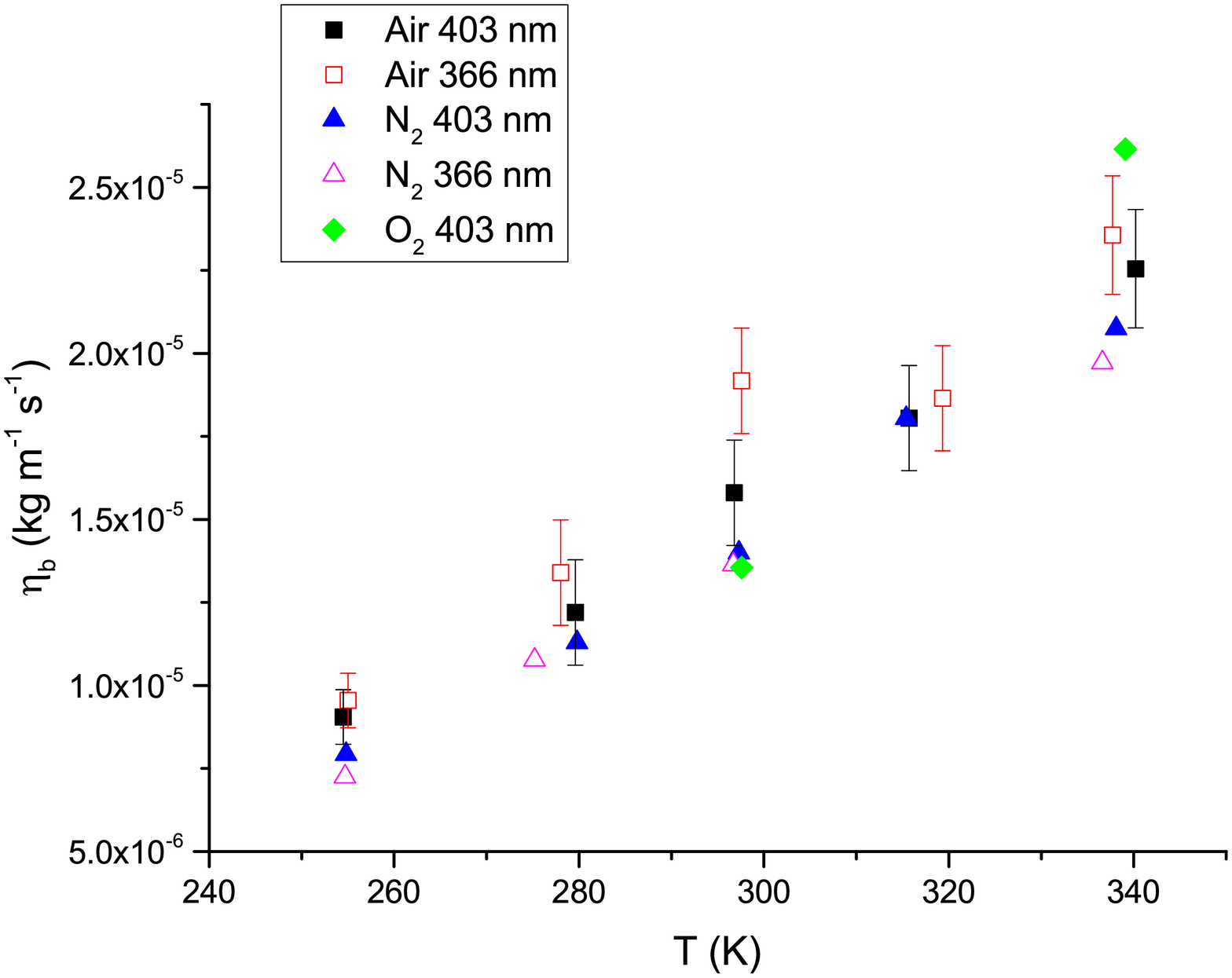}
\caption{\label{fig:eta_b_comparison} Results on bulk viscosity values $\eta_b$ as derived from Rayleigh-Brillouin scattering for air, N$_2$ and O$_2$ derived from this experiment at 403 nm and previous experiments at 366 nm~\cite{ZiyuGu2013Air,ZiyuGu2013N2}. Only for the cases of RB-scattering in air the uncertainties are explicitly specified. In order to avoid congestion, the error bars for other measurements, being similar to those of air, are not shown.}
\end{figure}

Values of bulk viscosity for air, N$_2$, and O$_2$, derived by comparing the 3 bar measurements to the Tenti S6 model using Eq.~(\ref{equ:least-square fit}), are plotted in black, blue, and green points in Fig.~\ref{fig:eta_b_comparison}, respectively. A clear increasing trend of the bulk viscosity with respect to temperature is detected for all of the three species, which can be explained by the fact that at higher temperatures more degrees of freedom will participate in the internal motion of the molecules and the relaxation time of the internal motion are shorter as the collisions happen more frequently. Uncertainty of this determination includes statistical errors resulting from the noise in the measurements, $\pm 0.9^\circ$ scattering angle uncertainty, 0.5\% uncertainty in pressure reading, and 0.5 K uncertainty in temperatures. Since all the data are obtained from the same setup with the same systematic uncertainty and similar signal-to-noise ratio, the error margins for all the bulk viscosity determinations are in the same order of magnitude. $\eta_b$ values for air and N$_2$ are compared with the ones derived previously from a  366 nm RB-scattering setup. Good agreement, within 1$\sigma$ overlap, is found for both of the species, demonstrating that the bulk viscosities for air and N$_2$ are insensitive to the small change of hypersound frequency associated with the $\sim 40$ nm variation in scattering wavelength.

The temperature-dependency of the bulk viscosity can be empirically interpreted in terms of a linear function:
\begin{equation}
\label{equ:T-dependence_bulk_viscosity}
\eta_b = \eta_b^0  + \gamma \cdot T
\end{equation}
Fitting of Eq.~(\ref{equ:T-dependence_bulk_viscosity}) to the values at 403 nm, as shown in Fig.~\ref{fig:eta_b_comparison}, gives $\eta_b^0 = (-3.15 \pm 0.22) \times10^{-5}$ kgm$^{-1}$s$^{-1}$ and $\gamma = (1.58 \pm 0.07) \times10^{-7}$ kgm$^{-1}$s$^{-1}$K$^{-1}$ for air, and $\eta_b^0 = (-3.30 \pm 0.26) \times10^{-5}$ kgm$^{-1}$s$^{-1}$ and $\gamma = (1.59 \pm 0.09) \times10^{-7}$ kgm$^{-1}$s$^{-1}$K$^{-1}$ for N$_2$.  Values for oxygen are found to not differ significantly from those for air and N$_2$.
For N$_2$ we experimentally establish the ratio $\eta_b/\eta= 0.79$ at room temperature, where experiments at ultrasound frequencies had yielded $\eta_b/\eta= 0.73$~\cite{Prangsma1973}. The good agreement between these values means that for the case of N$_2$ ultrasound and hypersound measurements deliver the same result.

%The present data confirm that the bulk viscosity is pressure-independent for light-scattering experiments in diluted gases up to 3 bar~\cite{ZiyuGu2014CO2}.
Values of $\eta_b$ for the gases at lower pressures, which are used as input parameters for the S6 simulations and listed in Table~\ref{Tab:conditions}, are calculated according to Eq.~(\ref{equ:T-dependence_bulk_viscosity}) in combination with fitted values of the two coefficients, $\eta_b^0$ and $\gamma$, given above.
The good agreement between the measurements and the calculations at lower pressures (see Figs.~\ref{fig:air_1bar_403nm_all} to \ref{fig:O2_403nm_all}) constitute a validation of the pressure-independence of the bulk viscosity, while the overall match between the measurements and the model demonstrates that the Tenti S6 model is valid to the same accuracy at the 403 nm wavelength as at 366 nm. Only the resulting values for $\eta_b$ between the 403 nm and 366 nm data for air at 300 K are found to slightly deviate, but still agreeing within 1.5 $\sigma$ combined error margins.

The present study demonstrates that the values for the bulk viscosity $\eta_b$ for nitrogen, oxygen and air
agree within the stated error margins, and can hence be considered as effectively the same. Similarly,
the values for shear viscosity $\eta$ and thermal conductivity $\kappa$ are also similar, while the internal
specific heat capacity $c_{int}=1$ for these diatomic molecular species. This means that the macroscopic transport coefficients underlying the RB-scattering profiles for air closely resemble those of nitrogen and oxygen. Combined
with the fact that the molecular masses of the atmospheric constituents are very close
(28 u for N$_2$ and 32 u for O$_2$), yielding an effective mass of 29 u for air particles, this makes that the
RB-profiles of air closely resembles those of N$_2$ and O$_2$, and that the treatment of air as a single species
gas with effective particle mass and transport coefficients holds so well.

\section{Simulations of RB-scattering for LIDAR applications}

\begin{figure}
\includegraphics[width=8.3cm]{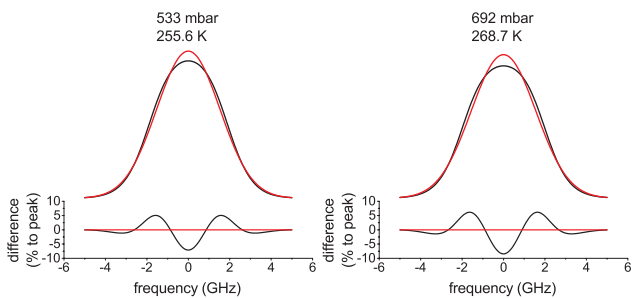}
\caption{\label{fig:lidar-prediction} Calculated spectral profiles of Rayleigh-Brillouin scattering for air in a typical LIDAR back-scattering ($180^{\circ}$) geometry probed with a 355 nm laser (black lines). The figure on the left represents the profile of air at $p=533$ mbar and $T=255.6$ K, the condition of 5 km altitude in the standard atmosphere, while the figure on the right represents the back-scattering profile at 3 km altitude level. The calculations are based on the Tenti S6 model with transport coefficients as calculated by Eqs.~(\ref{equ:shear_viscosity})-(\ref{equ:thermal_conductivity}) and the value for the bulk viscosity presently determined. For comparison purely Gaussian Doppler profiles for the same conditions are plotted (red lines). 7\% and 8.5\% differences between Gaussian and Tenti calculations are found.}
\end{figure}

With the $\eta_b$-values determined in the previous section and the other gas transport coefficients that can be obtained independently, the Tenti S6 model is proven to be valid within 2\% level in a wide range of temperatures, pressures and wavelengths including real atmospheric conditions. The spectral profiles and the comparison with model calculations, shown in Figs.~\ref{fig:air_1bar_403nm_all} to \ref{fig:O2_403nm_all}, serve as an illustration.
Hence, this model can be directly applied to LIDAR applications for exploring the properties of the atmosphere, such as wind speed profile retrieval or local temperature measurements. Scattering profiles of air at atmospheric conditions in the U.S. Standard Atmosphere model for 180$^\circ$ scattering angle and 355 nm, the often used scattering geometry and wavelength in LIDAR applications, are simulated and compared with purely Gaussian profiles in Fig.~\ref{fig:lidar-prediction}. Here, some typical conditions of $p=533$ mbar and $T=255.6$ K, corresponding to an altitude of 5 km in the Standard Atmosphere model, and of $p=692$ mbar and $T=268.7$ K, corresponding 3 km altitude, are chosen. Fig.~\ref{fig:lidar-prediction} clearly indicates that even at a height of 5 km in the atmosphere, a simple Gaussian assumption of the scattering profile would result in a 7\% error in the peak intensity. Such deviations had been discussed in studies preparing for the ADM-Aeolus wind LIDAR satellite mission of the European Space Agency, and have now been detailed and further quantified~\cite{Dabas2008,Pfaffrath2009}.
The Tenti S6 model, invoking the temperature-dependent values for the macroscopic gas transport coefficients including the presently derived values for the bulk viscosity provide a much better algorithm for atmospheric retrieval procedures.

\section{Conclusion}

A comprehensive study on spontaneous Rayleigh-Brillouin scattering in diatomic gases N$_2$ and O$_2$, and in air is reported. A large number of measurements recorded under different temperature-pressure conditions are compared with calculations based on the Tenti S6 model, yielding good agreement within 2\% of peak level intensities. Values for the bulk viscosity are determined at 403 nm and compared with the ones obtained with a 366 nm setup~\cite{ZiyuGu2013Air,ZiyuGu2013N2}, demonstrating that a slight change in hypersound frequency, associated with a wavelength change of $\sim40$ nm, does not affect the values of $\eta_b$.  

An important conclusion of the present study is that the approximation of air as a single component species
with effective transport coefficients and particle mass of 29 u holds well. This is understood from the fact that the
gas transport coefficients, bulk viscosity, shear viscosity, thermal conductivity and internal specific heat capacity
are all very much the same for air, N$_2$ and O$_2$, while the particle masses are also very similar for the three
species. In addition this explains why the Tenti S6 model, developed for single component species, is so well
applicable to air. 

The 2\% deviations between the measurements and the calculations are presently not understood. They may derive from effects of scattering due to temperature fluctuations at constant density, which were not considered in the Tenti S6 model, or be associated with the Wang-Chang and Uhlenbeck linearization, which is a fundamental approximation to derive the Tenti S6 model. Effects of rotational Raman scattering are not likely to have an influence on the RB scattering profile, since it has been filtered out by the 1 nm bandwidth filter. It would be interesting to further investigate depolarization effects induced by Raman scattering.

This study, together with previous ones reported by our group for a different wavelength, verifies that the Tenti S6 model, with the values for the temperature-dependent bulk viscosities determined, is an appropriate basis for atmospheric LIDAR application studies based on Rayleigh-Brillouin scattering.

\section*{Acknowledgments}
This work has been funded by the European Space Agency (ESA) contract ESTEC-21396/07/NL/HE-CCN-2. The authors thank A. G. Straume and O. Le Rille (ESA) for fruitful discussions. The code for computing the Tenti S6 model was provided by X. Pan and further modified by W. van de Water. The authors would like to thank B. Witschas and the German Aerospace Center (DLR) for providing the 1 nm bandwidth filter used for background removal.

%\bibliography{Ziyu_ref_list}

\providecommand{\noopsort}[1]{}\providecommand{\singleletter}[1]{#1}%

\end{document}